\documentclass[prb,twocolumn,showpacs]{revtex4-1}
\usepackage{graphicx}
\usepackage{amsfonts}
\usepackage{amssymb}
\usepackage{color}

\begin{document}

\title{Topological surface states in paramagnetic and antiferromagnetic
iron pnictides}

\author{Alexander Lau}
\affiliation{Institute of Theoretical Physics, Technische Universit\"at
Dresden, 01062 Dresden, Germany}

\author{Carsten Timm}
\email{carsten.timm@tu-dresden.de}
\affiliation{Institute of Theoretical Physics, Technische Universit\"at
Dresden, 01062 Dresden, Germany}

\date{\today}

\begin{abstract}
The electronic structure of iron pnictides is topologically nontrivial, leading
to the appearance of Dirac cones in the band structure for the antiferromagnetic
phase. Motivated by the analogy with Dirac cones in graphene, we explore the
possible existence
of topologically protected surface states. Surprisingly, bands of surface states
exist even in the paramagnetic state. A realistic five-orbital model
predicts two such bands. In the antiferromagnetic phase, these surface bands
survive but split. We obtain the bulk and surface dispersion from exact
diagonalization of two- and five-orbital models in a strip geometry and discuss
the results based on topology.
\end{abstract}

\pacs{
74.70.Xa, 
73.20.At, 
03.65.Vf, 
75.30.Fv  
}

\maketitle

\section{Introduction}

Two of the most active areas in condensed
matter physics concern the iron pnictides\cite{Joh10,DHD12} and topological
properties of matter.\cite{HaK10,QiZ11} The iron pnictides are of interest
since they show multiband superconductivity with high transition temperatures
competing with itinerant antiferromagnetism. Topology in condensed matter has
received a lot of attention in part because nontrivial topology can induce
surface or edge states. This paper is concerned with the
topological properties and surface states of iron pnictides.

In iron pnictides, several iron $3d$ orbitals contribute significantly
to the electronic states close to the Fermi energy. Ran \textit{et
al.}\cite{RWZ09} have realized that this multiorbital character can lead to
nontrivial topological properties. They find band touchings at the $(0,0)$ and
$(\pi,\pi)$ points in the Brillouin zone (BZ), which are associated with winding
numbers in orbital space. As a consequence, the formation of a spin-density wave
(SDW) with ordering vector $\mathbf{Q}=(\pi,0)$ cannot open a full gap but
rather leaves an even number of Dirac points.\cite{RWZ09} This holds
both for a two-orbital model including only the most important
$3d_{YZ}$ and $3d_{ZX}$ orbitals\cite{RQL08} and for a model including all
five $3d$ orbitals.\cite{KOA08} While the two-orbital model does not
give a realistic description,
it allows a simpler discussion of the topological properties. On the other hand,
the five-orbital model\cite{KOA08} gives a good account of the low-energy band
structure of the prototypical compound $\mathrm{LaFeAsO}$ and also leads to
reasonable predictions for the magnetic order vs.\ doping.\cite{BDT11,SBT12}
Evidence for the Dirac points has been
obtained from quantum-oscillation experiments,\cite{HaS09}
magnetotransport\cite{PBC13} and angle-resolved
photoemission (ARPES).\cite{RNS10,HTT11} However, conflicting evidence is
presented in \onlinecite{TKT11}.

Dirac points in the band structure have attracted a lot of attention in the
context of graphene.\cite{CGP09} There, Dirac points emerge due to the
two-sublattice structure of the honeycomb lattice. They are accompanied by
dispersing edge states at so-called zigzag edges,\cite{FWN96,RyH02,CGP06,CGP09}
which appear in the 1D edge BZ between the projections of the Dirac points.
Regardless of the different origins of the Dirac points, one may ask whether
they have similar consequences in pnictides and in graphene. Specifically,
do surface states also exist in the SDW phase of the iron pnictides?
Yang and Kee\cite{YaK10} have indeed found surface bands for a two-orbital
model with broken symmetry. However, the required combined orbital, SDW, and
charge-density-wave order is not realized in iron pnictides and actually opens a
full gap without Dirac points.

We will show that dispersing bands of surface states exist even in the
\emph{paramagnetic} phase of the iron pnictides, due to its topological
character. In the SDW phase, the surface
bands split. This is found for both two-orbital and five-orbital models but the
latter features two surface bands, whereas the former has only one. We will
explain the topological origin of the surface states.

\section{Models and method}

We follow Refs.\ \onlinecite{KOA08,RWZ09} in choosing
$3d$ orbitals with respect to the $X$ and $Y$ axes of the
tetragonal lattice, which are rotated by $45^\circ$ relative to the $x$ and $y$
axes of the square iron lattice. The two-orbital model\cite{RQL08,RWZ09}
retains only these two orbitals in a single-iron unit cell. The Hamiltonian is
$H=H_0+H_I$, where the non-interacting part for the extended system reads
$H_0 = \sum_{\mathbf{k}\sigma} \sum_{a,b=1}^2
  d^\dagger_{\mathbf{k}a\sigma} \mathcal{H}_{ab}(\mathbf{k})
  d_{\mathbf{k}b\sigma}$.
Here, $\mathcal{H}(\mathbf{k})$ is a $2\times2$ matrix in orbital space,
\begin{eqnarray}
\lefteqn{ \mathcal{H}(\mathbf{k}) = 2t_1\, (\cos k_x - \cos k_y)\, \tau^1 }
  \nonumber \\
&& {}- 2(t_2-t_2')\,\sin k_x \sin k_y\, \tau^3 \nonumber \\
&& {}+ [2(t_2+t_2')\,\cos k_x \cos k_y
  + 2t_1'\,(\cos k_x + \cos k_y)]\, \tau^0 ,\quad
\label{2_orbital_H0_1}
\end{eqnarray}
where $\tau^1$, $\tau^2$, $\tau^3$ are Pauli matrices, $\tau^0$ is the unit
matrix, and the orbital index $1$ corresponds to
$3d_{ZX}$ and $2$ to $3d_{YZ}$. We adopt the hopping parameters\cite{RWZ09}
$t_1=0.30\,\mathrm{eV}$, $t_1'=0.06\,\mathrm{eV}$,
$t_2=0.51\,\mathrm{eV}$, and $t_2'=0.09\,\mathrm{eV}$.

For the interactions we take\cite{RWZ09}
\begin{eqnarray}
H_I & = & \frac{U}{2} \sum_{\mathbf{i}} (\hat n_{\mathbf{i}1}^2 +
  \hat n_{\mathbf{i}2}^2)
  + (U-2J)\sum_{\mathbf{i}} n_{\mathbf{i}1}n_{\mathbf{i}2} \nonumber \\
&& {}+ J \sum_{\mathbf{i}} \sum_{\sigma\sigma'}
  d_{\mathbf{i}1\sigma}^{\dagger} d_{\mathbf{i}2\sigma'}^{\dagger}
  d_{\mathbf{i}1\sigma'} d_{\mathbf{i}2\sigma} \nonumber \\
&& {}+ J \sum_{\mathbf{i}} (d_{\mathbf{i}1\uparrow}^{\dagger}
  d_{\mathbf{i}1\downarrow}^{\dagger}d_{\mathbf{i}2\downarrow}
  d_{\mathbf{i}2\uparrow} + \mathrm{H.c.}) ,
\label{2_orbital_HI}
\end{eqnarray}
where $\hat n_{\mathbf{i}a} \equiv \sum_\sigma
d_{\mathbf{i}a\sigma}^{\dagger} d_{\mathbf{i}a\sigma}$, and the interaction
parameters are chosen as $U=1.20\,\mathrm{eV}$ and $J=0.12\,\mathrm{eV}$.
Assuming a SDW with ordering vector $\mathbf{Q}=(\pi,0)$
and spins pointing along the $S_z$ axis, a mean-field decoupling with
$\langle d_{\mathbf{i}a\sigma}^{\dagger} d_{\mathbf{i}b\sigma}\rangle =
  n_{ab} + (-1)^{i_x}\, \frac{\sigma}{2}\, m_{ab}$
gives the mean-field Hamiltonian
\begin{equation}
H_{\mathrm{MF}} = H_0 + \sum_{\mathbf{i}ab} (-1)^{i_x}M_{ab}
  \big(d_{\mathbf{i}a\uparrow}^{\dagger} d_{\mathbf{i}b\uparrow}
  - d_{\mathbf{i}a\downarrow}^{\dagger} d_{\mathbf{i}b\downarrow}\big) ,
\label{2_orbital_HMF}
\end{equation}
with $M_{11} = -(Um_{11}+Jm_{22})/2$, $M_{22} = -(Um_{22}+Jm_{11})/2$,
$M_{12} = M_{21} = -Jm_{21}$. The Hartree shifts $n_{ab}$ have been
absorbed into $H_0$. The parameters $m_{ab}$ are calculated self-consistently,
assuming half filling.

Edge states will be studied for a strip of width $W$ with (10) edges. Since the
strip is extended along the $y$ axis, $k_y$ is a good quantum number and
we carry out a Fourier transformation in the $y$ direction,
$d_{\mathbf{i}a\sigma} = N_y^{-1/2} \sum_{k_y} e^{ik_yi_y} d_{i_xk_ya\sigma}$.
The mean-field Hamiltonian then consists of blocks of dimensions $2W\times 2W$
for fixed $k_y$, $\sigma$. These blocks are
diagonalized numerically, giving the energy bands of the strip system. We here
assume that the SDW in the strip is described by the same order
parameters $m_{ab}$ as for the extended system.

The five-orbital model\cite{KOA08} includes all
hopping amplitudes larger than $10\,\mathrm{meV}$ up to fifth neighbors. The
hopping amplitudes and onsite energies are
obtained from density-functional calculations and are tabulated in Ref.\
\onlinecite{KOA08}. The interaction $H_I$ is analogous to
the two-orbital model, except that the interorbital terms in Eq.\
(\ref{2_orbital_HI}) now become
sums over all pairs of orbitals $a$, $b$, with $a<b$,
where $a,b=1,\ldots,5$ correspond to $3d_{3Z^2-R^2}$, $3d_{ZX}$, $3d_{YZ}$,
$3d_{X^2-Y^2}$, $3d_{XY}$.
The interaction parameters are chosen to be $U=1.0\,\mathrm{eV}$ and
$J=0.2\,\mathrm{eV}$.\cite{RWZ09}
Applying a mean-field decoupling as above, we obtain the
mean-field Hamiltonian (\ref{2_orbital_HMF}), where now the orbital indices
traverse $a,b=1,\ldots,5$, and
$M_{aa}=-(Um_{aa}+J\sum_{b\neq a}m_{bb})/2$ and
$M_{ab}= M_{ba} = -Jm_{ab}$ for $a\neq b$.
The mean-field parameters $m_{ab}$ are determined self-consistently, assuming
$6$ electrons per iron, corresponding to zero doping.
For the (10) strip, we proceed analogously to the two-orbital case. By
means of a Fourier transformation in the $y$ direction, the problem is reduced
to the diagonalization of $5W\times 5W$ Hamiltonian matrices for fixed $k_y$,
$\sigma$.

\begin{figure}[tb]
\centering
\includegraphics[width=\columnwidth]{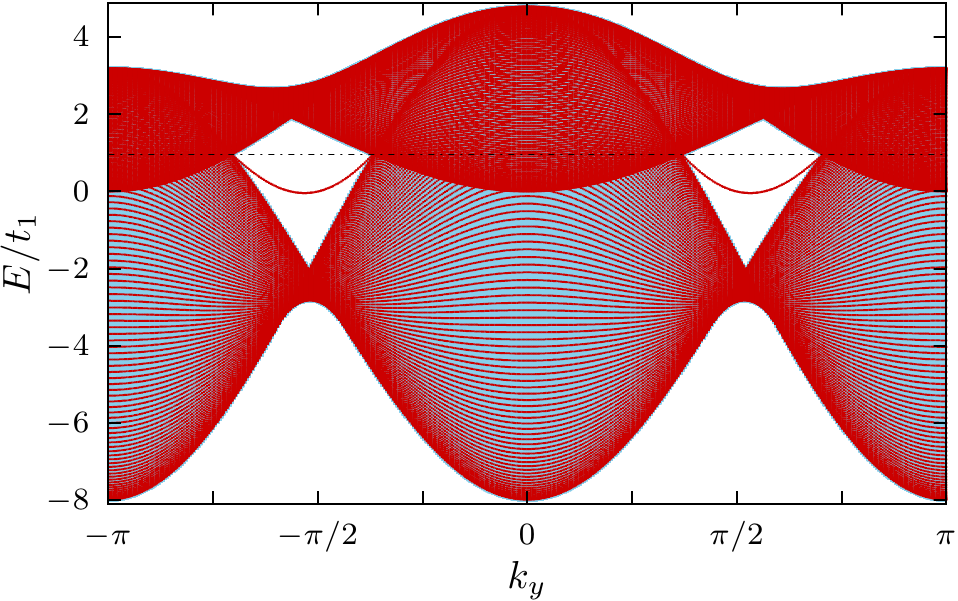}
\caption{(Color online) Energy bands of the two-orbital model in the absence of
SDW order:
Bands of a (10) strip of width $W=100$ (medium red/gray lines) compared to the
bulk bands projected onto the 1D BZ for the strip (light blue/gray region).
Note the bands of edge states for the strip.}
\label{fig:2_orbital_energies_nonmagnetic}
\end{figure}

\section{Results and discussion}

First, we consider the two-orbital model
without SDW order, i.e., with $m_{ab}=0$, in a (10) strip geometry. In
Fig.~\ref{fig:2_orbital_energies_nonmagnetic}, we compare
the energy bands of the (10) strip of width $W=100$ to the bands of the extended
system projected onto the (10) edge BZ. All energies are twofold spin
degenerate.
We see that the majority of the bands of the strip lie within the projected
continuum of bulk bands. However, there is an additional band that does not
agree with the bulk continuum. We have found that the corresponding states are
localized at the edges of the strip. A closer
look reveals that this band, ignoring spin degeneracy, actually comprises two
bands, which are only approximately degenerate. For finite widths $W$, the two
states at given $k_y$ correspond to bonding and antibonding combinations of
states localized at the two edges. Thus, there is a finite splitting, invisible
in Fig.~\ref{fig:2_orbital_energies_nonmagnetic}, which decreases exponentially
with increasing $W$.

\begin{figure}[tb]
\centering
\includegraphics[width=\columnwidth]{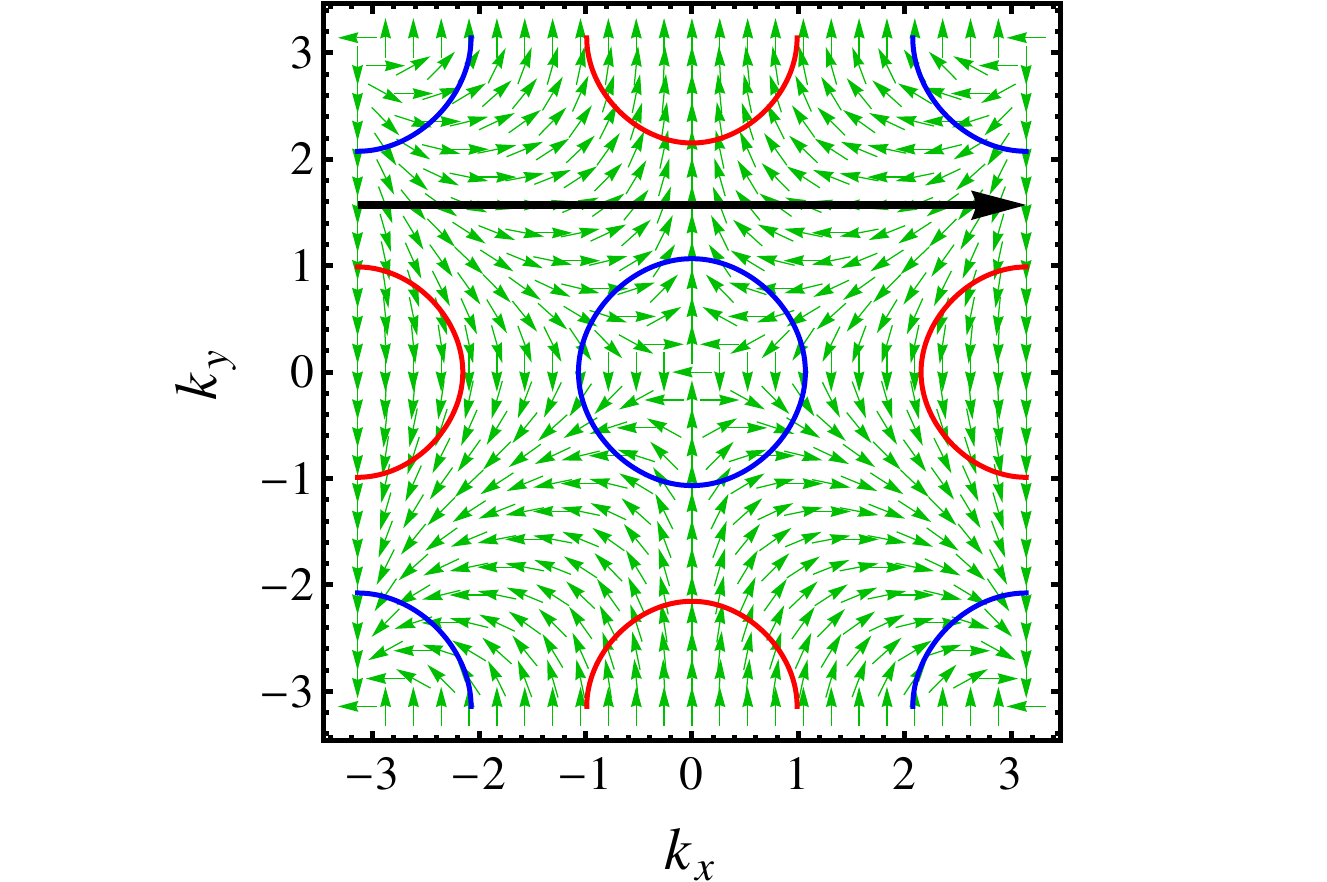}
\caption{(Color online) Fermi surfaces (solid lines) in the BZ
for the two-orbital model superimposed on the vector field
$(\cos\phi_\mathbf{k}, \sin\phi_\mathbf{k})$ (small arrows) showing
the nontrivial winding of the Hamiltonian $\mathcal{H}(\mathbf{k})$ in orbital
space. The black arrow denotes the BZ for an effective 1D model obtained by
fixing $k_y$, where the arrow is meant to indicate the projection onto the 1D
edge BZ.}
\label{fig:winding_2band}
\end{figure}

The origin of the edge states can be understood from a topological argument.
We start from the noninteracting first-quantized Hamiltonian
$\mathcal{H}(\mathbf{k})$ of Eq.\ (\ref{2_orbital_H0_1}).
We consider one spin sector throughout and suppress the spin index.
For the strip, $k_y$ is a constant of motion. For each
fixed value of $k_y$, we obtain an effective 1D model. For an extended system,
the BZ of this 1D model is the subset with $k_y=\mathrm{const}$ of the 2D BZ.
If the 1D BZ does not cross any Fermi surface, the 1D model is gapped in the
bulk. One such 1D BZ is shown by the black arrow in Fig.\
\ref{fig:winding_2band}. $\mathcal{H}(\mathbf{k})$ can be written
as\cite{RWZ09}
\begin{equation}
\mathcal{H}(\mathbf{k}) = a(\mathbf{k})\, \tau^0 + b(\mathbf{k})\,
  (\sin\phi_\mathbf{k}\, \tau^1 + \cos\phi_\mathbf{k}\, \tau^3) ,
\label{3.Hk.2}
\end{equation}
with $b(\mathbf{k})\ge 0$. The vector field $(\cos\phi_\mathbf{k},
\sin\phi_\mathbf{k})$ is plotted in Fig.\ \ref{fig:winding_2band}.
It exhibits vortices with vorticities $\pm 2$ at the band-touching points
$\mathbf{k}=(0,0)$ and $(\pi,\pi)$. Furthermore, a winding
number of $+1$ ($-1$) is aquired when moving from $k_x=-\pi$ to $k_x=+\pi$ along
a line with constant $k_y>0$ ($k_y<0$). The effective 1D system
thus has a nontrivial topological structure in orbital space.\cite{RWZ09}

The appearance of edge states is best understood by deforming the
Hamiltonian of the effective 1D system into one with a topologically protected
winding number.
For fixed $k_y>0$ ($k_y<0$ is analogous), we deform the Hamiltonian
into $\tilde \mathcal{H}(k_x) = \cos k_x\, \tau^1 - \sin k_x\, \tau^3$.
This deformation does not change the topology of the bands and, in particular,
leaves the energy gap open. $\tilde \mathcal{H}(k_x)$ is unitarily equivalent to
\begin{equation}
\hat \mathcal{H}(k_x) \equiv e^{-i \frac{\pi}{4} \tau^1}
  \tilde \mathcal{H}(k_x)\, e^{i \frac{\pi}{4} \tau^1}
  = \left(\begin{array}{cc}
  0 & e^{-ik_x} \\
  e^{ik_x} & 0
  \end{array}\right) .
\label{3.Hflat.2}
\end{equation}
In the bulk, this new Hamiltonian has the eigenvalues $\pm 1$ for every $k_x$,
i.e., it has flat bands. Furthermore, $\hat \mathcal{H}(k_x)$ is time-reversal
symmetric. The antiunitary time-reversal operator $T$ can be written as
$T=KU_T$, where $K$ is the complex conjugation in our basis and $U_T$ is unitary
and must satisfy $U_T \hat \mathcal{H}^*(-k_x) U_T^\dagger = \hat
\mathcal{H}(k_x)$, which is fulfilled for $U_T=\tau^0$. Note that $T$ squares to
$T^2 = \tau^0$. $\hat \mathcal{H}(k_x)$ is also charge-conjugation
symmetric. The antiunitary charge-conjugation operator $C$ can be written as
$C=KU_C$, where $U_C$ is unitary and must satisfy $U_C \hat \mathcal{H}^*(-k_x)
U_C^\dagger = - \hat \mathcal{H}(k_x)$, which is fulfilled for $U_C=\tau^3$. We
find $C^2 = K\tau^3 K\tau^3 = \tau^0$. These symmetry properties imply that
the deformed 1D model belongs to the Altland-Zirnbauer class BDI.\cite{Zir96}
This class allows a $\mathbb{Z}$ topological invariant in 1D.\cite{SRF08}

For the strip, the effective 1D model is a finite chain of length $W$. The
$\mathbb{Z}$ topological invariant means that states localized
at the ends can exist, but does not guarantee that $\hat \mathcal{H}(k_x)$
has any. However, this is easily seen by transforming $\hat
\mathcal{H}(k_x)$ into real space, which gives
$\hat H = \sum_{j=1}^{W-1} ( d_{j+1,\uparrow}^\dagger d_{j,\downarrow}
  + d_{j,\downarrow}^\dagger d_{j+1,\uparrow} )$,
where ${\uparrow}$, ${\downarrow}$ now refer to the \emph{orbital} pseudospin.
The ${\uparrow}$ (${\downarrow}$) state at $j=1$
($j=W$) is not coupled to any other state. Thus there are two zero-energy states
localized at the ends. These arguments can be made for any $k_y$ for which the
1D BZ does not intersect a Fermi surface.

What does this tell us about the original Hamiltonian $\mathcal{H}(\mathbf{k})$
with fixed $k_y$? That Hamiltonian has neither $T$ nor $C$ symmetry and is thus
in class A, which is topologically trivial in 1D. Our point is that the
original Hamiltonian can be obtained from $\hat \mathcal{H}(k_x)$ by a
continuous deformation without closing the gap. During the deformation, the $T$
and $C$ symmetries are lost so that the zero-energy end states are no longer
protected. However, the energy of the end states evolves
smoothly during the deformation. Hence, for $\mathcal{H}(\mathbf{k})$ there are
still two edge states for every $k_y$ for which the effective 1D Hamiltonian is
gapped. These states have no reason to be at zero energy and will generally have
an exponentially small overlap with each other, which splits their energies.
In principle, the edge states are not protected against merging with the bulk
continuum. However, we find separate edge states wherever the bulk is gapped,
presumably due to level repulsion between edge and bulk states.
Note the similarity to graphene: The edge band at graphene zigzag edges is a
flat zero-energy band only for a model without next-nearest-neighbor hopping. In
real graphene, it is dispersing.\cite{CGP06,CGP09}

\begin{figure}[tb]
\centering
\includegraphics[width=\columnwidth]{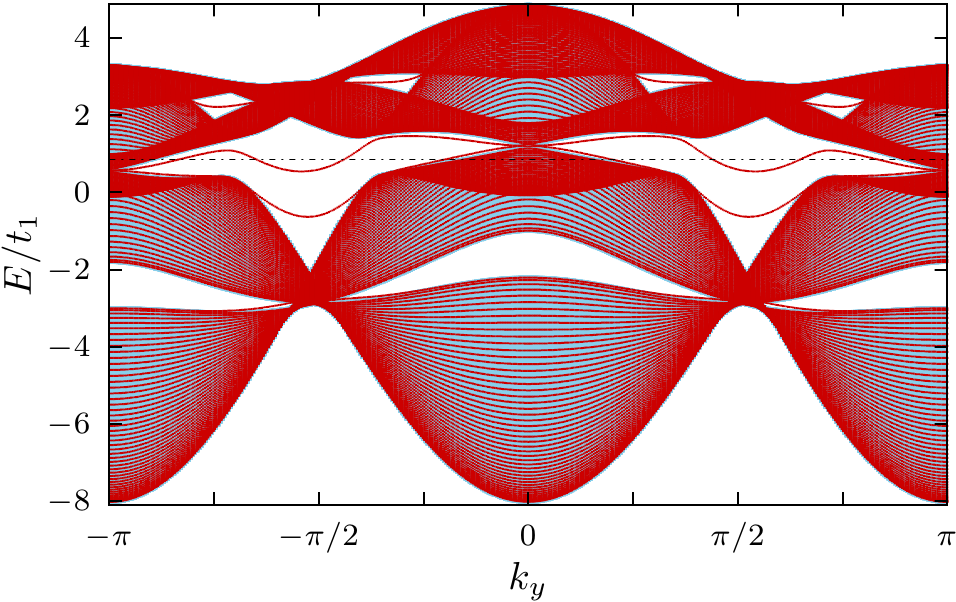}
\caption{(Color online) Energy bands of the two-orbital model with SDW order:
Bands of a (10) strip of width $W=100$ (medium red/gray lines) compared to
the bulk bands projected onto the 1D BZ for the strip (light blue/gray region).}
\label{fig:2_orbital_energies_magnetic}
\end{figure}

In the presence of a SDW with ordering vector $\mathbf{Q}=(\pi,0)$, the unit
cell of the iron square lattice is doubled in the $x$ direction. Therefore, the
bands are folded into the magnetic BZ. SDW formation opens gaps at some of the
new band crossings but not at all of them---the bands still stick together at
Dirac points.\cite{RWZ09} Figure \ref{fig:2_orbital_energies_magnetic} shows
the band structure of the (10) strip compared to the projected bulk bands.
Spin degeneracy is not lifted by the SDW since the mean-field Hamiltonian is
invariant under combined spin rotation and spatial reflection $x\to-x$. However,
the near degeneracy between bonding and antibonding combinations of edge states
is strongly broken. We instead find two edge states per spin direction, which
are localized mainly at one edge and are split in energy due to the opposite
exchange field at the two edges.
Note that the edge bands are connected to the bulk bands at the
projected Dirac points, similar to graphene. Moreover,
Fig.~\ref{fig:2_orbital_energies_magnetic} shows that additional edge bands
appear within the new gaps away from the Fermi energy.

\begin{figure}[tb]
\centering
\includegraphics[width=\columnwidth]{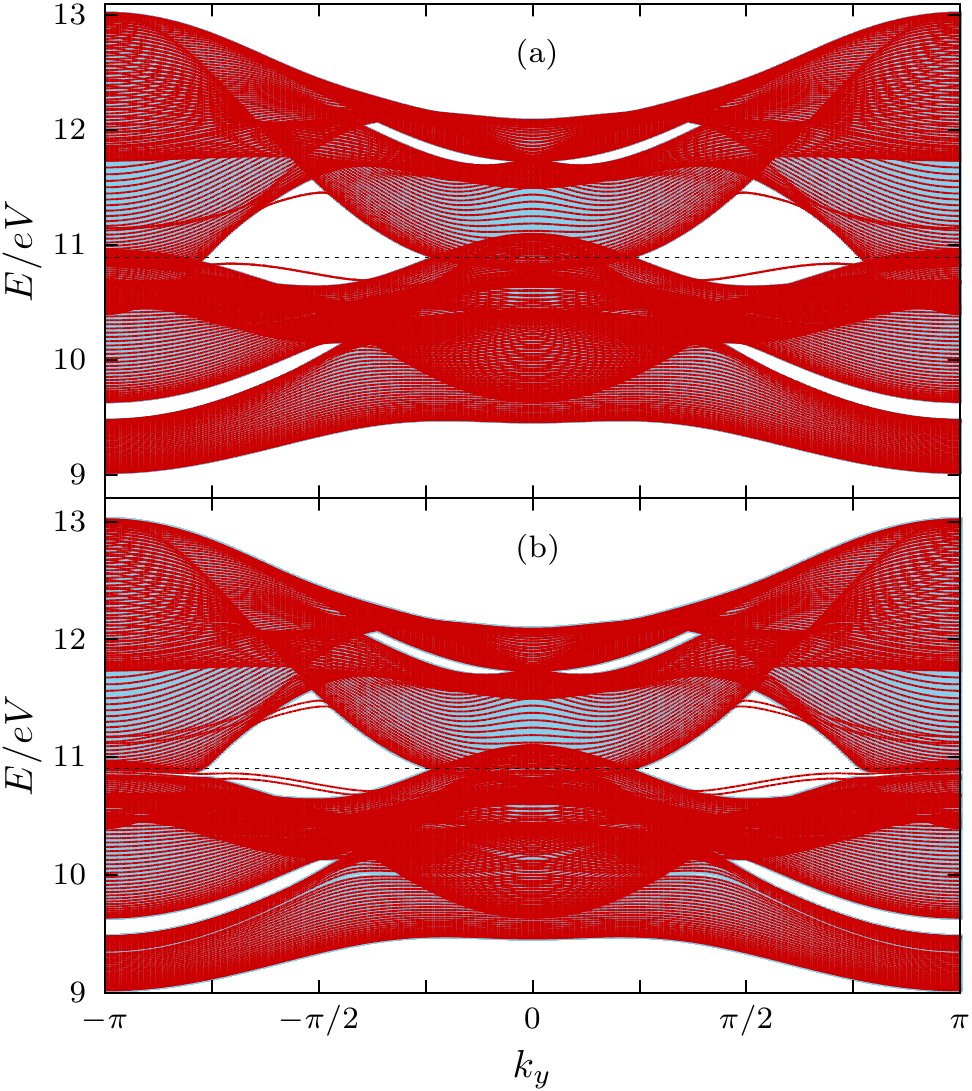}
\caption{(Color online) Energy bands of the five-orbital model (a) without and
(b) with SDW order: Bands of a (10) strip of width $W=60$ (medium red/gray
lines) compared to the bulk bands projected onto the 1D BZ for the strip (light
blue/gray regions).}
\label{fig:5_orbital_energies}
\end{figure}

To check whether the edge bands also occur for the five-orbital
model,\cite{KOA08,RWZ09} we plot in Fig.\ \ref{fig:5_orbital_energies}
the energy bands of the (10) strip compared to the projected bulk bands
for the paramagnetic and the SDW phase. In the paramagnetic phase we now find
two edge bands, each of which is twofold spin degenerate and, in the wide
limit, also twofold degenerate between bonding and antibonding
combinations. SDW order again leads to the opening of gaps at
crossings of folded bands. The twofold spin degeneracy remains but the
asymptotic degeneracy between bonding and antibonding states is lifted, for the
same reason as for the two-orbital model.

To understand the appearance of surface states in the five-orbital model,
we again consider effective gapped 1D models obtained by fixing $k_y$ to values
for which the path through the BZ at constant $k_y$ does not intersect a
Fermi surface. In this case, two bands lie above the Fermi energy and
three below. The Hamiltonian can be deformed continuously
without closing the gap performing the following
steps. (i) The hopping amplitudes between the $d_{3Z^2-R^2}$ and all other
orbitals are tuned to zero and the resulting decoupled band is
shifted down in energy, effectively removing it from the model, together with
two electrons per iron. (ii) The hopping amplitudes beyond next-nearest
neighbors are tuned to zero. All components of the
Hamiltonian $\mathcal{H}(k_x)$ can now
be written as linear combinations of $\cos k_x$, $\sin k_x$, and a constant.
(iii) All constant terms are tuned to zero. (iv) The coefficients of $\sin k_x$
in the diagonal components are tuned to zero. (v) The coefficients of $\cos
k_x$ in the off-diagonal components are tuned to zero. (vi) The remaining
coefficients in diagonal components are tuned to unity and in off-diagonal
components to $1/\sqrt{2}$. A unitary transformation in orbital space maps the
resulting Hamiltonian onto
\begin{equation}
\hat \mathcal{H}(k_x) = \left(\begin{array}{cccc}
  0 & e^{-ik_x} & 0 & 0 \\
  e^{ik_x} & 0 & 0 & 0 \\
  0 & 0 & 0 & e^{-ik_x} \\
  0 & 0 & e^{ik_x} & 0
  \end{array}\right) .
\end{equation}
Since this deformed Hamiltonian just consists of two copies of the Hamiltonian
in Eq.\ (\ref{3.Hflat.2}), it has two sets of zero-energy end states of distinct
orbital character. Upon deforming the Hamiltonian back into the original one,
the end states for every considered $k_y$ continuously develop into dispersing
bands. In addition, the degeneracy between the edge states of different orbital
character is lifted so that we end up with two bands.

\section{Conclusions}

We have shown that two models of iron pnictides predict
the existence of dispersing bands of surface states at (100) surfaces for the
paramagnetic state. They could be used to probe the topologically nontrivial
electronic structure. These bands are twofold spin degenerate and have an
additional twofold degeneracy in the limit of a thick slab due to the decoupling
of the states localised at the two surfaces. The main difference between the
two-orbital and the five-orbital model is that the latter predicts two instead
of one such bands. Their existence can be understood from a topological argument
based on the continuous deformation of the Hamiltonian into one in
Altland-Zirnbauer class BDI. In the SDW state, where we had guessed from an
analogy to graphene that surface states might exist, the surface bands split
into two twofold degenerate bands. It should be noted that while the
two models inherit the surface states from topologically nontrivial
Hamiltonians, they are not themselves topologically nontrivial. Hence, the
surface states are not robust against disorder scattering. The analogy is thus
more with graphene than with topological insulators.

The origin of the surface states discussed
here is completely different from surface states at (001) surfaces of iron
pnictides of the 1111 family, such as $\mathrm{LaOFeAs}$,
which have been observed by ARPES.\cite{LYM09,ELK10} In this case, surface bands
result from the polar nature of these surfaces.\cite{ELK10} Surface states of
this kind are not expected for $\mathrm{LiFeAs}$ and $\mathrm{NaFeAs}$ since the
(001) surface of these compounds is not polar and they are indeed not seen by
ARPES.\cite{LKB10,YLM12}

The surface states predicted here could be detected by ARPES or scanning
tunneling spectroscopy on (100) surfaces, which are however challenging to
prepare. An alternative could be tunneling into the edges of thin
(001) films, either
using an STM tip, a technique that has been successful for the
detection of edge states in graphene,\cite{KFE05}
or one of the setups discussed in Ref.\ \onlinecite{Sei11} in
the context of Josephson junctions. Such
tunneling experiments would probe the density of states, which is expected to
be enhanced close to (100) edges of the film. In particular,
the van Hove singularities associated with extrema of the surface bands lead to
an enhancement of the density of states since the surface states are
essentially one-dimensional. It is an interesting
question what happens to the surface states when superconductivity sets in.
Superconductivity of the $s^{+-}$ type preferred for iron pnictides could
introduce energy gaps in the surface bands discussed here but also
induce additional surface-bound states.\cite{OnT09} Both would also be relevant
for recent suggestions to engineer topological superconductors by using the
proximity effect of superconducting iron pnictides.\cite{ZKM13}

\section*{Acknowledgments}

We thank P. M. R. Brydon, M. Daghofer, and A. P. Schnyder for
helpful discussions. Support by the Deutsche Forschungsgemeinschaft through
Research Training School GRK 1621 and Priority Programme SPP 1458 is
acknowledged.


\begin{thebibliography}{99}

\bibitem{Joh10}D. Johnston, Adv.\ Phys.\ \textbf{59}, 803 (2010).

\bibitem{DHD12}P. Dai, J. Hu, and E. Dagotto, Nature Phys.\ \textbf{8}, 709
(2012).

\bibitem{HaK10}M. Z. Hasan and C. L. Kane, Rev.\ Mod.\ Phys.\ \textbf{82}, 3045
(2010).

\bibitem{QiZ11}X.-L. Qi and S.-C. Zhang, Rev.\ Mod.\ Phys.\ \textbf{83}, 1057
(2011).

\bibitem{RWZ09}Y. Ran, F. Wang, H. Zhai, A. Vishwanath, and D.-H. Lee, Phys.\
Rev.\ B \textbf{79}, 014505 (2009).

\bibitem{RQL08}S. Raghu, X.-L. Qi, C.-X. Liu, D. J. Scalapino, and S.-C. Zhang,
Phys.\ Rev.\ B \textbf{77}, 220503(R) (2008).

\bibitem{KOA08}K. Kuroki, S. Onari, R. Arita, H. Usui, Y. Tanaka, H. Kontani,
and H. Aoki, Phys.\ Rev.\ Lett.\ \textbf{101}, 087004 (2008).

\bibitem{BDT11}P. M. R. Brydon, M. Daghofer, and C. Timm, J. Phys.: Condens.\
Matter \textbf{23}, 246001 (2011).

\bibitem{SBT12}J. Schmiedt, P. M. R. Brydon, and C. Timm, Phys.\ Rev.\ B
\textbf{85}, 214425 (2012).

\bibitem{HaS09}N. Harrison and S. E. Sebastian, Phys.\ Rev.\ B \textbf{80},
224512 (2009); M. Sutherland, D. J. Hills, B. S. Tan, M. M. Altarawneh, N.
Harrison, J. Gillett, E. C. T. O'Farrell, T. M. Benseman, I. Kokanovic, P.
Syers, J. R. Cooper, and S. E. Sebastian, Phys.\ Rev.\ B \textbf{84}, 180506(R)
(2011).

\bibitem{PBC13}I. Pallecchi, F. Bernardini, F. Caglieris, A. Palenzona, S.
Massidda, and M. Putti, Eur.\ Phys.\ J. B \textbf{86}, 338 (2013).

\bibitem{RNS10}P. Richard, K. Nakayama, T. Sato, M. Neupane, Y.-M. Xu, J. H.
Bowen, G. F. Chen, J. L. Luo, N. L. Wang, X. Dai, Z. Fang, H. Ding, and T.
Takahashi, Phys.\ Rev.\ Lett.\ \textbf{104}, 137001 (2010).

\bibitem{HTT11}K. K. Huynh, Y. Tanabe, and K. Tanigaki, Phys.\ Rev.\ Lett.\
\textbf{106}, 217004 (2011).

\bibitem{TKT11}T. Terashima, N. Kurita, M. Tomita, K. Kihou, C.-H. Lee, Y.
Tomioka, T. Ito, A. Iyo, H. Eisaki, T. Liang, M. Nakajima, S. Ishida, S.
Uchida, H. Harima, and S. Uji, Phys.\ Rev.\ Lett.\ \textbf{107}, 176402 (2011).

\bibitem{CGP09} A. H. Castro Neto, F. Guinea, N. M. R. Peres, K. S. Novoselov,
and A. K. Geim, Rev.\ Mod.\ Phys.\ \textbf{81}, 109 (2009).

\bibitem{FWN96}M. Fujita, K. Wakabayashi, K. Nakada, and K. Kusakabe, J. Phys.\
Soc.\ Jpn.\ \textbf{65}, 1920 (1996); K. Nakada, M. Fujita, G. Dresselhaus, and
M. S. Dresselhaus, Phys.\ Rev.\ B \textbf{54}, 17954 (1996).

\bibitem{CGP06}A. H. Castro Neto, F. Guinea, and N. M. R. Peres, Phys.\ Rev.\ B
\textbf{73}, 205408 (2006).

\bibitem{RyH02}S. Ryu and Y. Hatsugai, Phys.\ Rev.\ Lett.\ \textbf{89}, 077002
(2002).

\bibitem{YaK10}B.-J. Yang and H.-Y. Kee, Phys.\ Rev.\ B \textbf{82}, 195126
(2010), in particular note Fig.~7.

\bibitem{Zir96}M. R. Zirnbauer, J. Math.\ Phys.\ \textbf{37}, 4986 (1996);
A. Altland and M. R. Zirnbauer, Phys.\ Rev.\ B \textbf{55}, 1142 (1997).

\bibitem{SRF08}A. P. Schnyder, S. Ryu, A. Furusaki, and A. W. W. Ludwig, Phys.\
Rev.\ B \textbf{78}, 195125 (2008).

\bibitem{LYM09}D. H. Lu, M. Yi, S.-K. Mo, J. G. Analytis, J.-H. Chu,
A. S. Erickson, D. J. Singh, Z. Hussain, T. H. Geballe, I. R. Fisher, and Z.-X.
Shen, Physica C \textbf{469}, 452 (2009).

\bibitem{ELK10}H. Eschrig, A. Lankau, and K. Koepernik, Phys.\ Rev.\ B
\textbf{81}, 155447 (2010).

\bibitem{LKB10}A. Lankau, K. Koepernik, S. Borisenko, V. Zabolotnyy,
B. B\"uchner, J. van den Brink, and H. Eschrig, Phys.\ Rev.\ B \textbf{82},
184518 (2010).

\bibitem{YLM12}M. Yi, D. H. Lu, R. G. Moore, K. Kihou, C.-H. Lee, A.
Iyo, H. Eisaki, T. Yoshida, A. Fujimori, and Z.-X. Shen, New J. Phys.\
\textbf{14}, 073019 (2012).

\bibitem{KFE05}Y. Kobayashi, K.-i. Fukui, T. Enoki, K. Kusakabe, and
Y. Kaburagi, Phys.\ Rev.\ B \textbf{71}, 193406 (2005);
Y. Kobayashi, K.-i. Fukui, T. Enoki, and K. Kusakabe, Phys.\ Rev.\ B
\textbf{73}, 125415 (2006).

\bibitem{Sei11}P. Seidel, Supercond.\ Sci.\ Technol.\ \textbf{24}, 043001
(2011).

\bibitem{OnT09}S. Onari and Y. Tanaka, Phys.\ Rev.\ B \textbf{79},
174526 (2009).

\bibitem{ZKM13}F. Zhang, C. L. Kane, and E. J. Mele, Phys.\ Rev.\ Lett.\
\textbf{111}, 056402 (2013).





\end{thebibliography}
\end{document}